\documentclass[pre,amsmath,amssymb,reprint]{revtex4-1}
\usepackage{epsfig}                                                                 
\usepackage{dcolumn}                                                                
\usepackage{bm}                                                                     
\usepackage[pdfstartview=FitH,bookmarks=false]{hyperref}                            

\begin{document}
\title[Peculiar property of noble gases and the Enskog--Vlasov model]{Peculiar property of noble gases\\and its explanation through the
Enskog--Vlasov model}
\author{E. S. Benilov}
 \altaffiliation[]{Department of Mathematics and Statistics, University of Limerick, V94 T9PX, Ireland}
 \email{Eugene.Benilov@ul.ie}
 \homepage{http://www.staff.ul.ie/eugenebenilov/hpage/}
\author{M. S. Benilov}
 \altaffiliation[]{Departamento de F\'{\i}sica, CCCEE, Universidade da Madeira, Largo do
Munic\'{\i}pio, 9000 Funchal, Portugal; Instituto de Plasmas e Fus\~{a}o Nuclear,
Instituto Superior T\'{e}cnico, Universidade de Lisboa, Portugal}
 \email{benilov@uma.pt}
 \homepage{http://fisica.uma.pt/ingles/pessoal/Mikhail_Benilov/}
\date{\today}

\begin{abstract}
A new observation is presented that the ratio of the critical and
triple-points temperatures, $T_{cr}/T_{tp}$, of neon, argon, krypton, and
xenon fit within a narrow interval: $T_{cr}/T_{tp}=1.803\pm0.5\%$, and the
same applies to the density ratio: $n_{cr}/n_{tp}=$ $0.3782\pm1.7\%$ (of the
two remaining noble gases, helium does not have a triple point and, for radon,
$n_{tp}$ is unknown). We explain this peculiar property by the fact that the
molecules of noble gases are nearly spherical, as a result of which they
satisfy the Enskog--Vlasov (EV) kinetic model based on the approximation of
hard spheres. The EV model has also allowed us to identify two more parameter
combinations which are virtually the same for all noble gases.
\end{abstract}
\maketitle

\section{Introduction}

Noble gases have been studied for decades -- and yet no-one has observed that
their parameters form certain non-dimensional `invariants', which hardly
change from gas to gas. In the present paper, we identify four such invariants
and explain their universality using the Enskog--Vlasov (EV) model.

The EV kinetic equation comprises the Enskog collision integral for dense
fluids \cite{Enskog22} and a Vlasov term describing the van-der-Waals force
(similar to that for the electromagnetic force in plasma \cite{Vlasov68}). The
first version of the EV model \cite{Desobrino67,Grmela71} was based on the
original form of the Enskog integral -- which, as shown in Ref.
\cite{VanbeijerenErnst73a}, does not comply with the Onsager relations. Ref.
\cite{VanbeijerenErnst73a} has also proposed a modification of the Enskog
integral that is free from this shortcoming, which was incorporated in the EV
model in Refs. \cite{KarkheckStell81,StellKarkheckVanbeijeren83}. A
restriction of the coefficients of the Enskog--Vlasov equation guaranteeing
that it satisfies an H-theorem has been formulated in Refs.
\cite{GrmelaGarciacolin80,GrmelaGarciacolin80b,Grmela81}, and Ref.
\cite{BenilovBenilov18} proposed a version of the EV equation that does
satisfy this restriction and conserves energy as well (all of the previous
versions did not).\medskip

\emph{The EV model}---The Enskog integral is based on an approximation of the
molecules by hard spheres -- hence, the Enskog--Vlasov model should be best
applicable to noble fluids, whose molecules are nearly spherical (e.g., they
have neither dipole nor quadrupole asymmetry). It has been tested for neon,
argon, krypton, and xenon \cite{BenilovBenilov18}, and it was shown that the
thermodynamic properties of these fluids are indeed consistent with the
constraints implied by the EV model. In particular, the per-molecule internal
energy $U_{pm}$ and entropy $S_{pm}$ are reasonably accurately described by%
\begin{equation}
U_{pm}=\frac{3}{2}k_{B}T-\frac{1}{2}En, \label{1}%
\end{equation}%
\begin{equation}
S_{pm}=k_{B}\left[  \ln\frac{T^{3/2}}{n}-\Theta(D^{3}n)\right]  , \label{2}%
\end{equation}
where $k_{B}$ is the Boltzmann constant, $T$ is the temperature, $n$ is the
number density, $D$ is the effective molecular diameter, the Vlasov parameter
$E$ characterizes the strength of the van-der-Waals force. The function
$\Theta(\xi)$ describes the non-ideal part of the fluid's entropy and, thus,
vanishes at zero density,%
\begin{equation}
\Theta(0)=0, \label{3}%
\end{equation}
whereas the EV model implies \cite{BenilovBenilov18} that%
\begin{equation}
\Theta^{\prime}(0)=\frac{2\pi}{3}, \label{4}%
\end{equation}
where $\Theta^{\prime}(\xi)=\mathrm{d}Q/\mathrm{d}\xi$. As shown in Ref.
\cite{BenilovBenilov18}, Eqs. (\ref{1})--(\ref{2}) correspond to the following
equation of state (EoS):%
\begin{equation}
p=nk_{B}T\left[  1+D^{3}n\,\Theta^{\prime}(D^{3}n)\right]  -\frac{1}{2}En^{2},
\label{5}%
\end{equation}
where $p$ is the pressure, and the following expression for the per-molecule
Gibbs free energy:%
\begin{multline}
G_{pm}=k_{B}T\left[  \ln(nT^{-3/2})\right. \\
+\left.  \Theta(D^{3}n)+D^{3}n\,\Theta^{\prime}(D^{3}%
n)\vphantom{T^{-3/2}}\right]  -En, \label{6}%
\end{multline}
Before using the EV model, one should calibrate it, i.e. fix $E$, $D$, and
$\Theta(\xi)$. Note that the first two parameters are specific to the fluid,
whereas $\Theta(\xi)$ is supposed to be a universal function characterizing
all noble fluids.

The Vlasov parameter $E$ was determined for each of the four fluids under
consideration by fitting a linear dependence to the empiric data
\cite{LinstromMallard97} for $U-\tfrac{3}{2}k_{B}T$ on the critical isobar, as
a function of $n$ (see Table \ref{table1}) \footnote{In Ref.
\cite{BenilovBenilov18}, $E$ was calculated using the isobar with the pressure
being twice the critical pressure. The difference between that result and the
present one is less than 0.5\%, which proves the robust nature of the employed
method for determining $E$.}. Table \ref{table1} also presents the parameters
of the critical and triple points of the four fluids \cite{LinstromMallard97}
to be used later.

\begin{table*}[ptb]
\centering
\begin{ruledtabular}
\begin{tabular}[c]{|p{0.6cm}||p{2.7cm}||p{2.5cm}|p{1.5cm}|p{1.5cm}||p{2.5cm}|p{1.5cm}|p{1.5cm}|}
&\vspace{-0.5mm} $E N_{A}^2~(\mathrm{J~l/mol} ^2)$\vspace{1.5mm}%
&\vspace{-0.5mm} $n_{tp} / N_{A}~(\mathrm{mol/l})$ &\vspace{-0.5mm}\hspace{-0.1mm} $T_{tp}~(\mathrm{K})$ &\vspace{-0.5mm} $p_{tp}~(\mathrm{bar})$%
&\vspace{-0.5mm} $n_{cr} / N_{A}~(\mathrm{mol/l})$ &\vspace{-0.5mm}\hspace{-0.1mm} $T_{cr}~(\mathrm{K})$ &\vspace{-0.5mm} $p_{cr}~(\mathrm{bar})$\\
\hline\hline
\vspace{-0.5mm}\hspace{0.15mm} $\mathrm{Ne}$\vspace{1.5mm} &\vspace{-0.5mm}\hspace{10.1mm} $51.6$ &\vspace{-0.5mm}\hspace{5.5mm} $62.059$%
&\vspace{-0.5mm}\hspace{1.4mm} $24.562$ &\vspace{-0.5mm}\hspace{-0.1mm} $0.43464$%
&\vspace{-0.5mm}\hspace{5.5mm} $23.882$ &\vspace{-0.5mm}\hspace{1.3mm} $44.4918$ &\vspace{-0.5mm}\hspace{0.3mm} $26.786$\\
\hline
\vspace{-0.5mm}\hspace{0.15mm} $\mathrm{Ar}$\vspace{1.5mm} &\vspace{-0.5mm}\hspace{8.5mm} $325$ &\vspace{-0.5mm}\hspace{5.5mm} $35.465$%
&\vspace{-0.5mm}\hspace{1.4mm} $83.8058$ &\vspace{-0.5mm}\hspace{-0.1mm} $0.68891$%
&\vspace{-0.5mm}\hspace{5.5mm} $13.4074$ &\vspace{-0.5mm}\hspace{-0.1mm} $150.687$ &\vspace{-0.5mm}\hspace{0.3mm} $48.630$\\
\hline
\vspace{-0.5mm}\hspace{0.15mm} $\mathrm{Kr}$\vspace{1.5mm} &\vspace{-0.5mm}\hspace{8.5mm} $550$ &\vspace{-0.5mm}\hspace{5.5mm} $29.197$%
&\vspace{-0.5mm}\hspace{-0.2mm} $115.77$ &\vspace{-0.5mm}\hspace{-0.1mm} $0.73503$%
&\vspace{-0.5mm}\hspace{5.5mm} $10.85$ &\vspace{-0.5mm}\hspace{-0.1mm} $209.48$ &\vspace{-0.5mm}\hspace{0.3mm} $55.250$\\
\hline
\vspace{-0.5mm}\hspace{0.15mm} $\mathrm{Xe}$\vspace{1.5mm} &\vspace{-0.5mm}\hspace{8.5mm} $983$ &\vspace{-0.5mm}\hspace{5.5mm} $22.592$%
&\vspace{-0.5mm}\hspace{-0.2mm} $161.4$ &\vspace{-0.5mm}\hspace{-0.1mm} $0.81748$%
&\vspace{-0.5mm}\hspace{7.1mm} $8.4$ &\vspace{-0.5mm}\hspace{-0.1mm} $289.733$ &\vspace{-0.5mm}\hspace{0.3mm} $58.420$\\
\end{tabular}
\end{ruledtabular}
\caption{The dimensional parameters of neon, argon, krypton, and xenon. The
Vlasov parameter $E$, $n_{tp}$, and $n_{cr}$ are normalized using the Avogadro
constant $N_{A}$.}%
\label{table1}%
\end{table*}

As shown in Ref. \cite{BenilovBenilov18}, the effective molecular diameter $D$
can be related to the triple-point density $n_{tp}$,%
\begin{equation}
D=n_{tp}^{-1/3}, \label{7}%
\end{equation}
whereas the function $\Theta(\xi)$ will be discussed later.

\section{The main result}

Introduce the following nondimensional parameters:%
\begin{equation}
\alpha=\frac{T_{cr}}{T_{tp}},\qquad\beta=\frac{n_{cr}}{n_{tp}}, \label{8}%
\end{equation}%
\begin{equation}
\gamma=\dfrac{k_{B}T_{tp}}{En_{tp}},\qquad\delta=\dfrac{p_{cr}}{k_{B}%
T_{cr}n_{cr}}. \label{9}%
\end{equation}
If one of these parameters is calculated using the data from Table
\ref{table1} for the four noble fluids under consideration, the resulting four
values (see Table \ref{table2}) fit into a fairly narrow interval:%
\[
\alpha=1.803\pm0.5\%,\hspace{1.15cm}\beta=0.3782\pm1.7\%,
\]%
\[
\gamma=0.06186\pm3.1\%,\qquad\delta=0.2959\pm2.4\%.
\]
If neon is excluded, the universal nature of invariants (\ref{8})--(\ref{9})
becomes a little more evident:%
\[
\alpha=1.802\pm0.4\%,\hspace{1.15cm}\beta=0.3748\pm0.9\%,
\]%
\[
\gamma=0.06010\pm0.4\%,\qquad\delta=0.2905\pm0.6\%.
\]
The fact that neon is slightly off can be explained by the fact that its
triple-point temperature is considerably lower than those of the other four
fluids; as a result, its liquid phase may be influenced by quantum effects.
Note that the van-der-Waals EoS yields for $\delta$ a universal value of
$0.375$, which however differs significantly from that observed for noble gases.

\begin{table}[ptb]
\centering
\begin{ruledtabular}
\begin{tabular}[c]{|p{0.6cm}||p{1.5cm}|p{1.5cm}|p{2cm}|p{2cm}|}
&\vspace{-0.5mm}\hspace{3.5mm} $\dfrac{T_{cr}}{T_{tp}}$\vspace{1.5mm} &\vspace{-0.5mm}\hspace{3.2mm} $\dfrac{n_{cr}}{n_{tp}}$%
&\vspace{-0.5mm}\hspace{4.1mm} $\dfrac{k_{B} T_{tp}}{E n_{tp}}$ &\vspace{-0.5mm}\hspace{2.1mm} $\dfrac{p_{cr}}{k_{B} T_{cr} n_{cr}}$\\
\hline\hline
\vspace{-0.5mm}\hspace{0.15mm} $\mathrm{Ne}$\vspace{1.5mm} &\vspace{-0.5mm}\hspace{1.5mm} $1.8114$ &\vspace{-0.5mm}\hspace{1mm} $0.38483$%
&\vspace{-0.5mm}\hspace{3mm} $0.063774$ &\vspace{-0.5mm}\hspace{3.1mm} $0.30320$\\
\hline
\vspace{-0.5mm}\hspace{0.15mm} $\mathrm{Ar}$\vspace{1.5mm} &\vspace{-0.5mm}\hspace{1.5mm} $1.7980$ &\vspace{-0.5mm}\hspace{1mm} $0.37805$%
&\vspace{-0.5mm}\hspace{3mm} $0.060454$ &\vspace{-0.5mm}\hspace{3.1mm} $0.28950$\\
\hline
\vspace{-0.5mm}\hspace{0.15mm} $\mathrm{Kr}$\vspace{1.5mm} &\vspace{-0.5mm}\hspace{1.5mm} $1.8094$ &\vspace{-0.5mm}\hspace{1mm} $0.37161$%
&\vspace{-0.5mm}\hspace{3mm} $0.059942$ &\vspace{-0.5mm}\hspace{3.1mm} $0.29237$\\
\hline
\vspace{-0.5mm}\hspace{0.15mm} $\mathrm{Xe}$\vspace{1.5mm} &\vspace{-0.5mm}\hspace{1.5mm} $1.7951$ &\vspace{-0.5mm}\hspace{1mm} $0.37181$%
&\vspace{-0.5mm}\hspace{3mm} $0.060427$ &\vspace{-0.5mm}\hspace{3.1mm} $0.28870$\\
\end{tabular}
\end{ruledtabular}
\caption{The nondimensional parameters of neon, argon, krypton, and xenon.}%
\label{table2}%
\end{table}

To explain the constancy of invariants (\ref{8})--(\ref{9}), assume that the
four fluids under consideration are described by the EV model with the same
function $\Theta(\xi)$. Recall also that a fluid's critical and triple points
both lie on the curve in the $\left(  n,T\right)  $ plane, representing the
relationship between the parameters of the saturated vapor and liquid. This
curve can be obtained through the Maxwell construction, i.e. by equating the
two phases' pressures and chemical potentials (for a single-component fluid,
the latter coincides with the per-molecule Gibbs free energy). Recalling thus
(\ref{5})--(\ref{6}) and introducing the non-dimensional variables%
\begin{equation}
\xi=D^{3}n,\qquad\tau=\frac{k_{B}D^{3}}{E}T, \label{10}%
\end{equation}
we obtain%
\begin{equation}
\xi_{v}\tau\left[  1+\xi_{v}\Theta^{\prime}(\xi_{v})\right]  -\frac{\xi
_{v}^{2}}{2}=\xi_{l}\tau\left[  1+\xi_{l}\Theta^{\prime}(\xi_{l})\right]
-\frac{\xi_{l}^{2}}{2}, \label{11}%
\end{equation}%
\begin{multline}
\tau\left[  \ln\xi_{v}+\Theta(\xi_{v})+\xi_{v}\Theta^{\prime}(\xi_{v})\right]
-\xi_{v}\\
=\tau\left[  \ln\xi_{l}+\Theta(\xi_{l})+\xi_{l}\Theta^{\prime}(\xi
_{l})\right]  -\xi_{l}, \label{12}%
\end{multline}
where the subscripts $_{v}$ and $_{l}$ mark the parameters of the vapor and
liquid phases, respectively.

Physically, vapor and liquid can coexist only if $T<T_{cr}$ -- hence, Eqs.
(\ref{11})--(\ref{12}) have a non-trivial solution ($\xi_{v}\neq\xi_{l}$) only
if $\tau<\tau_{cr}$, where $\tau_{cr}$ is the nondimensional $T_{cr}$.
Straightforward calculations show that $\tau_{cr}$ and the corresponding
nondimensional density $\xi_{cr}$ are determined by%
\begin{equation}
3\xi_{cr}^{2}\Theta^{\prime\prime}(\xi_{cr})+\xi_{cr}^{3}\Theta^{\prime
\prime\prime}(\xi_{cr})=1, \label{13}%
\end{equation}%
\begin{equation}
\tau_{cr}=\frac{\xi_{cr}}{1+2\xi_{cr}\Theta^{\prime}(\xi_{cr})+\xi_{cr}%
^{2}\Theta^{\prime\prime}(\xi_{cr})}. \label{14}%
\end{equation}
To find the parameters of the triple point, observe that calibration (\ref{7})
and nondimensionalization (\ref{10}) imply%
\begin{equation}
\left(  \xi_{l}\right)  _{tp}=1. \label{15}%
\end{equation}
Thus, the nondimensional triple-point temperature can be found by letting in
Eqs. (\ref{11})--(\ref{12}) $\xi_{l}=1$ and $\tau=\tau_{tp}$,%
\begin{equation}
\xi_{v}\tau_{tp}\left[  1+\xi_{v}\Theta^{\prime}(\xi_{v})\right]  -\frac
{\xi_{v}^{2}}{2}=\tau_{tp}\left[  1+\Theta^{\prime}(1)\right]  -\frac{1}{2},
\label{16}%
\end{equation}%
\begin{multline}
\tau_{tp}\left[  \ln\xi_{v}+\Theta(\xi_{v})+\xi_{v}\Theta^{\prime}(\xi
_{v})\right]  -\xi_{v}\\
=\tau_{tp}\left[  \Theta(1)+\Theta^{\prime}(1)\right]  -1. \label{17}%
\end{multline}
Eq. (\ref{16})--(\ref{17}) determine $\tau_{tp}$ and the corresponding
nondimensional density $\xi_{v}$ of the saturated vapor (in what follows, the
latter will not be needed).

Recalling nondimensionalization (\ref{10}), relationship (\ref{7}), the
latter's nondimensional equivalent (\ref{15}), and EoS (\ref{5}), we can
express invariants (\ref{8})--(\ref{9}) in the form%
\begin{equation}
\alpha=\frac{\tau_{cr}}{\tau_{tp}},\qquad\beta=\xi_{cr}, \label{18}%
\end{equation}%
\begin{equation}
\gamma=\tau_{tp},\qquad\delta=1+\Theta^{\prime}(1)-\frac{1}{2\tau_{tp}}.
\label{19}%
\end{equation}
Now, the universality of these parameters follows from the mere fact that Eqs.
(\ref{13})--(\ref{14}) and (\ref{16})--(\ref{17}) -- which determine $\xi
_{cr}$, $\tau_{cr}$, and $\tau_{tp}$ -- do not include any fluid-specific parameters.

Note that none of our conclusions derived so far depends on the specific form
of the function $\Theta(\xi)$, as long as it is the same for all four fluids
under consideration.

\section{Calibrating the EV model}

The invariants found turn out to be helpful for the inner workings of the EV
model, as they help us to finish calibrating it for noble fluids [by fixing
$\Theta(\xi)$].

We shall approximate $\Theta(\xi)$ by a fifth-degree polynomial. Given
restrictions (\ref{3})--(\ref{4}), this amounts to%
\begin{equation}
\Theta=\frac{2\pi}{3}\xi+a_{2}\xi^{2}+a_{3}\xi^{3}+a_{4}\xi^{4}+a_{5}\xi^{5}.
\label{20}%
\end{equation}
We have deduced the coefficients $a_{2,3,4,5}$ from the requirement that the
invariants $\alpha$, $\beta$, $\gamma$, and $\delta$ assume the correct
values, where the \textquotedblleft correct\textquotedblright\ means
\textquotedblleft the average over argon, krypton, and xenon\textquotedblright%
\ (to eliminate quantum effects -- no matter how weak they are -- neon was
excluded). To do so, we used Eqs. (\ref{13})--(\ref{14}), (\ref{16}%
)--(\ref{17}), and (\ref{18})--(\ref{20}) to relate $a_{2,3,4,5}$ to $\alpha$,
$\beta$, $\gamma$, and $\delta$ -- and thus obtained%
\begin{align}
a_{2}  &  =-1.8103,\qquad a_{3}=9.6325,\label{21}\\
a_{4}  &  =-12.831,\qquad a_{5}=6.2501. \label{22}%
\end{align}
Since the function $\Theta(\xi)$ is now known, we can calculate the parameters
of the saturated vapor and liquid from Eqs. (\ref{11})--(\ref{12}), and use
Eq. (\ref{5}) to find the EoS. These results have been compared to the
corresponding empiric data \cite{LinstromMallard97}.

\begin{figure}
\includegraphics[width=\columnwidth]{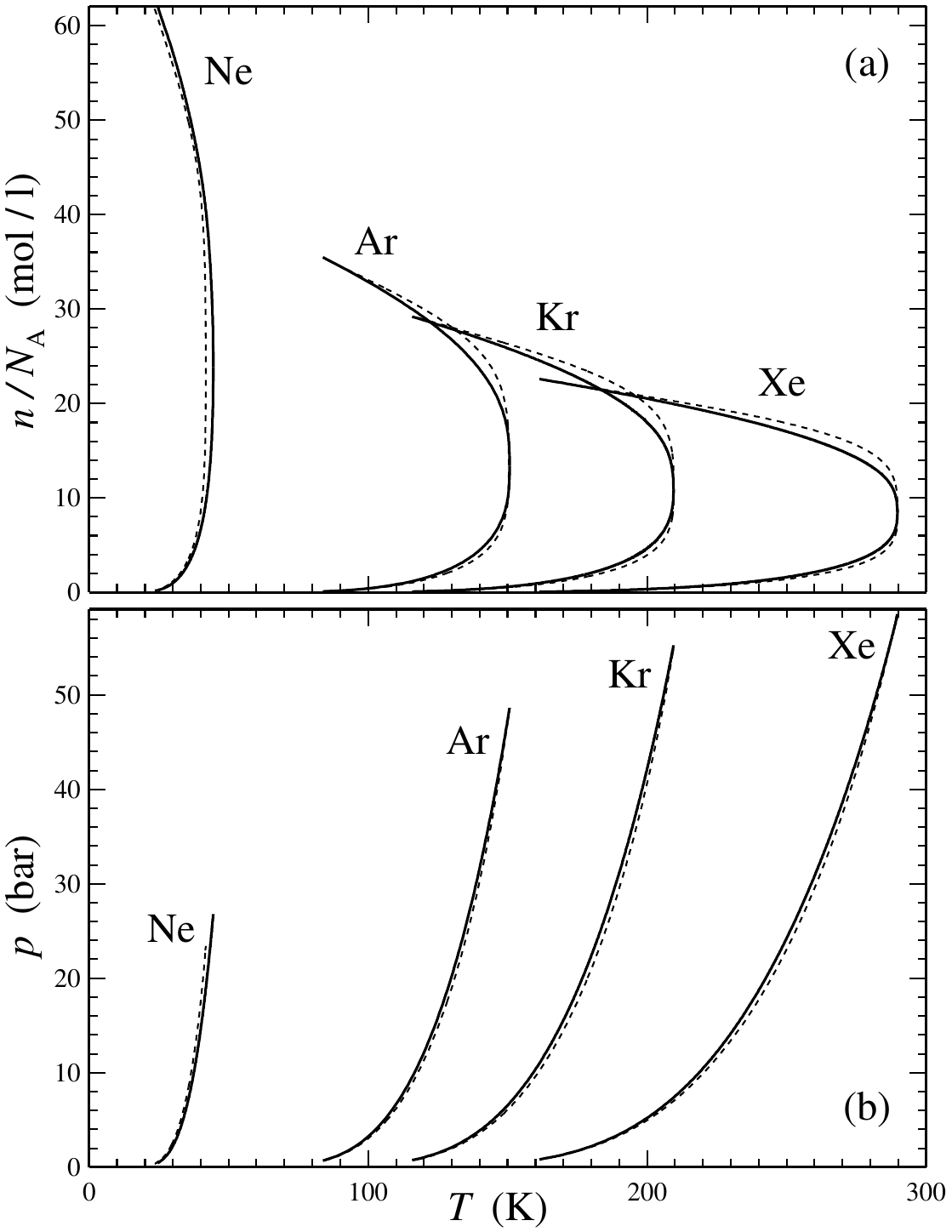}
\caption{The parameters of phase transitions of noble fluids. The solid curves show the empiric data \cite{LinstromMallard97} and the dotted ones, the results obtained through the EV\ model. (a) The molar densities of the saturated vapor and liquid (the upper and lower parts of the curves,
respectively) vs. $T$. (b) The pressure of the saturated vapor vs. $T$.}
\label{fig1}
\end{figure}

\begin{figure*}
\includegraphics[width=\columnwidth]{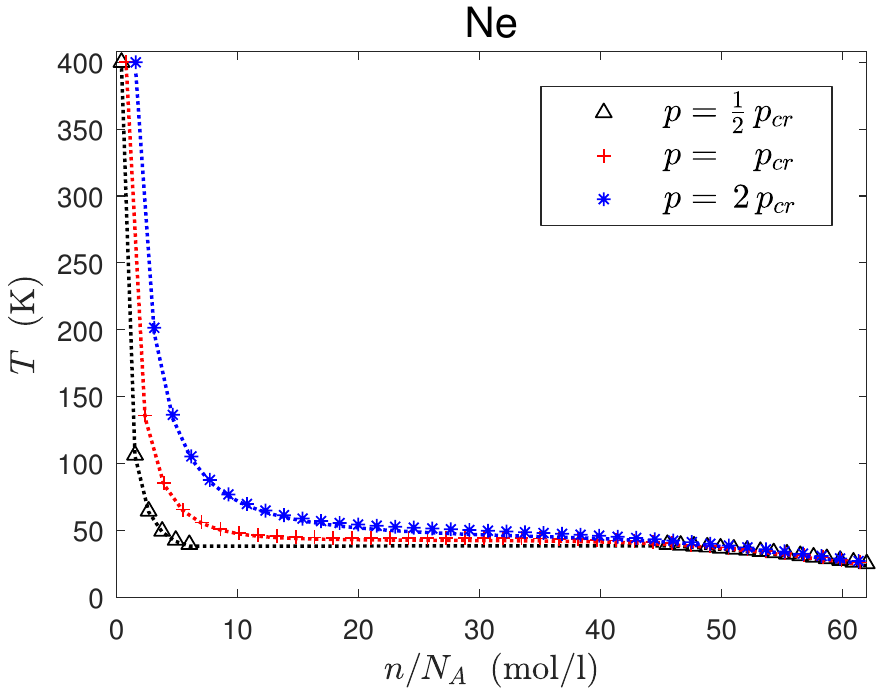}
\includegraphics[width=\columnwidth]{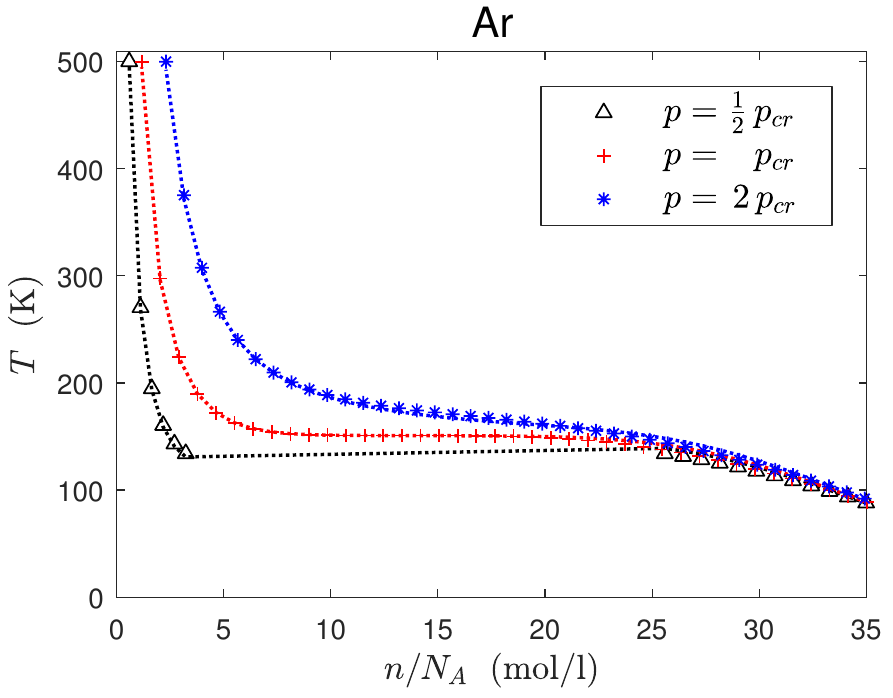}
\newline\newline
\includegraphics[width=\columnwidth]{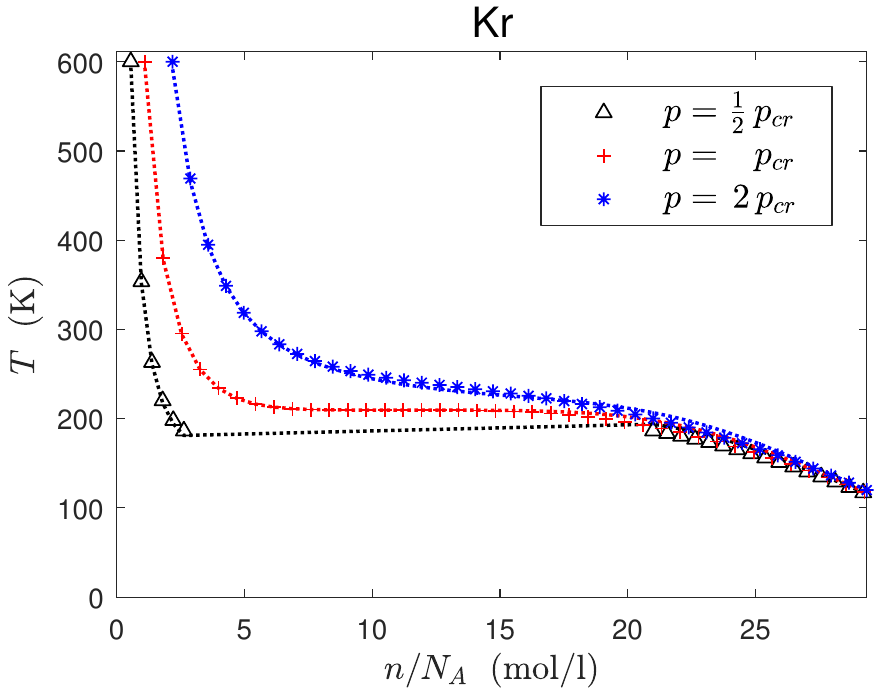}
\includegraphics[width=\columnwidth]{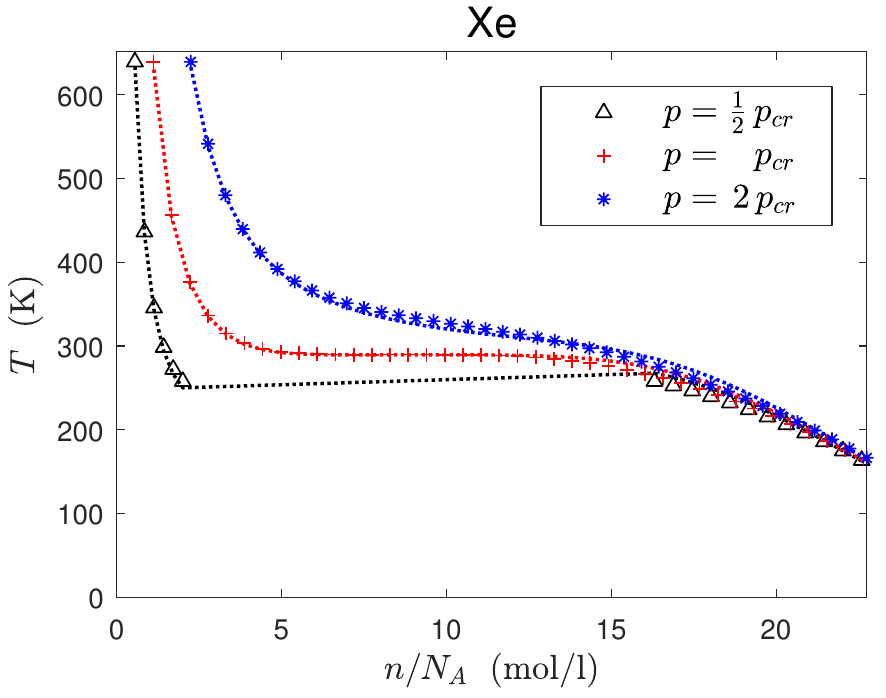}
\caption{A comparison of the Enskog--Vlasov EoS (\ref{5}), (\ref{7}), (\ref{20})--(\ref{22}) (dotted curves) and the empiric data \cite{LinstromMallard97} (non-connected symbols) for neon, argon, krypton, and xenon. The panel labels (circled) indicate the corresponding fluid.}
\label{fig2}
\end{figure*}

Since we have calibrated the EV model using the parameters of the critical and
triple points of argon, krypton, and xenon, it comes as no surprise that their
phase-transition properties are described well -- and even those of neon are
described reasonably accurately (see Fig. \ref{fig1}). More surprisingly, the
same calibration also provides an accurate description of the noble fluids'
EoS (see Fig. \ref{fig2}) \footnote{Note that the EoS-derived calibration
proposed in\ Ref. \cite{BenilovBenilov18} is sufficiently accurate only for
the EoS, but is less so for the parameters of the saturated vapor and liquid
(which we did not realize at the time).}.

\section{Summary and concluding remarks}

The main result of the present work is an observation that certain
characteristics of phase transitions of noble fluids hardly change from fluid
to fluid. Other groups of substances do not seem to have this peculiar
property, not even the halogens (which are the closest neighbors of noble
gases in the periodic table). Indeed, for fluorine, chlorine, bromine, and
iodine, $T_{cr}/T_{tp}=2.409\pm12.1\%$, i.e. the spread of this parameter is
considerably larger than that for noble fluids. We attribute this difference
to the fact that the halogen molecules are strongly asymmetric, so the
Enskog--Vlasov model does not describe them as accurately as it does noble fluids.

Note that the constancy of invariants (\ref{8})--(\ref{9}) can also be
explained using basic dimensional analysis and the assumption that, in the
fluid(s) under consideration, the potential of the van-der-Waals force
involves no more than 2 dimensional parameters -- such as, for example, the Lennard-Jones
potential truncated at the molecule's diameter. This approach was used in
Ref. \cite{RabinovichVassermanNedostupVeksler88} to explain constancy of the
(critical-point) parameter $\delta$, but it has never been applied to
triple-point characteristics, such as $\alpha$, $\beta$, and $\gamma$.
Furthermore, the EV model allows one to explain the constancy of all four
characteristics regardless of the number of parameters in the van-der-Waals force.

Finally, we emphasize that we do not advertise the EV model as a means of
predicting properties of fluids, as there are tools (e.g., Refs.
\cite{LinstromMallard97,TegelerSpanWagner99}) which do this job with a higher
accuracy. It should rather be used as a \emph{kinetic} equation, and the
accuracy of its thermodynamic predictions is just an indicator of its overall
accuracy. This is an important point, as several version of the Enskog--Vlasov
kinetic equation have been used for applications (see Refs.
\cite{FrezzottiBarbante17,FrezzottiGibelliLockerbySprittles18} and references therein).

\begin{acknowledgments}
We acknowledge the support of the Science Foundation Ireland through
grant 12/IA/1683, and that of Funda\c{c}\~{a}o para a Ci\^{e}ncia e a Tecnologia of
Portugal through project Pest-OE/UID/FIS/50010/2013.

We are also grateful to V. Yu. Zitserman for helpful remarks.\end{acknowledgments}


\begin{thebibliography}{17}%
\makeatletter
\providecommand \@ifxundefined [1]{%
 \@ifx{#1\undefined}
}%
\providecommand \@ifnum [1]{%
 \ifnum #1\expandafter \@firstoftwo
 \else \expandafter \@secondoftwo
 \fi
}%
\providecommand \@ifx [1]{%
 \ifx #1\expandafter \@firstoftwo
 \else \expandafter \@secondoftwo
 \fi
}%
\providecommand \natexlab [1]{#1}%
\providecommand \enquote  [1]{``#1''}%
\providecommand \bibnamefont  [1]{#1}%
\providecommand \bibfnamefont [1]{#1}%
\providecommand \citenamefont [1]{#1}%
\providecommand \href@noop [0]{\@secondoftwo}%
\providecommand \href [0]{\begingroup \@sanitize@url \@href}%
\providecommand \@href[1]{\@@startlink{#1}\@@href}%
\providecommand \@@href[1]{\endgroup#1\@@endlink}%
\providecommand \@sanitize@url [0]{\catcode `\\12\catcode `\$12\catcode
  `\&12\catcode `\#12\catcode `\^12\catcode `\_12\catcode `\%12\relax}%
\providecommand \@@startlink[1]{}%
\providecommand \@@endlink[0]{}%
\providecommand \url  [0]{\begingroup\@sanitize@url \@url }%
\providecommand \@url [1]{\endgroup\@href {#1}{\urlprefix }}%
\providecommand \urlprefix  [0]{URL }%
\providecommand \Eprint [0]{\href }%
\providecommand \doibase [0]{http://dx.doi.org/}%
\providecommand \selectlanguage [0]{\@gobble}%
\providecommand \bibinfo  [0]{\@secondoftwo}%
\providecommand \bibfield  [0]{\@secondoftwo}%
\providecommand \translation [1]{[#1]}%
\providecommand \BibitemOpen [0]{}%
\providecommand \bibitemStop [0]{}%
\providecommand \bibitemNoStop [0]{.\EOS\space}%
\providecommand \EOS [0]{\spacefactor3000\relax}%
\providecommand \BibitemShut  [1]{\csname bibitem#1\endcsname}%
\let\auto@bib@innerbib\@empty
\bibitem [{\citenamefont {Enskog}(1922)}]{Enskog22}%
  \BibitemOpen
  \bibfield  {author} {\bibinfo {author} {\bibfnamefont {D.}~\bibnamefont
  {Enskog}},\ }\href@noop {} {\bibfield  {journal} {\bibinfo  {journal} {Kungl.
  Svenska Vetenskaps Akad. Handl.}\ }\textbf {\bibinfo {volume} {63}},\
  \bibinfo {pages} {1} (\bibinfo {year} {1922})}\BibitemShut {NoStop}%
\bibitem [{\citenamefont {Vlasov}(1968)}]{Vlasov68}%
  \BibitemOpen
  \bibfield  {author} {\bibinfo {author} {\bibfnamefont {A.~A.}\ \bibnamefont
  {Vlasov}},\ }\href {\doibase 10.1070/pu1968v010n06abeh003709} {\bibfield
  {journal} {\bibinfo  {journal} {Sov. Phys. Usp.}\ }\textbf {\bibinfo {volume}
  {10}},\ \bibinfo {pages} {721} (\bibinfo {year} {1968})}\BibitemShut
  {NoStop}%
\bibitem [{\citenamefont {de~Sobrino}(1967)}]{Desobrino67}%
  \BibitemOpen
  \bibfield  {author} {\bibinfo {author} {\bibfnamefont {L.}~\bibnamefont
  {de~Sobrino}},\ }\href {\doibase 10.1139/p67-035} {\bibfield  {journal}
  {\bibinfo  {journal} {Can. J. Phys.}\ }\textbf {\bibinfo {volume} {45}},\
  \bibinfo {pages} {363} (\bibinfo {year} {1967})}\BibitemShut {NoStop}%
\bibitem [{\citenamefont {Grmela}(1971)}]{Grmela71}%
  \BibitemOpen
  \bibfield  {author} {\bibinfo {author} {\bibfnamefont {M.}~\bibnamefont
  {Grmela}},\ }\href {\doibase 10.1007/BF01011389} {\bibfield  {journal}
  {\bibinfo  {journal} {J. Stat. Phys.}\ }\textbf {\bibinfo {volume} {3}},\
  \bibinfo {pages} {347} (\bibinfo {year} {1971})}\BibitemShut {NoStop}%
\bibitem [{\citenamefont {van Beijeren}\ and\ \citenamefont
  {Ernst}(1973)}]{VanbeijerenErnst73a}%
  \BibitemOpen
  \bibfield  {author} {\bibinfo {author} {\bibfnamefont {H.}~\bibnamefont {van
  Beijeren}}\ and\ \bibinfo {author} {\bibfnamefont {M.~H.}\ \bibnamefont
  {Ernst}},\ }\href {\doibase 10.1016/0031-8914(73)90372-8} {\bibfield
  {journal} {\bibinfo  {journal} {Physica}\ }\textbf {\bibinfo {volume} {68}},\
  \bibinfo {pages} {437} (\bibinfo {year} {1973})}\BibitemShut {NoStop}%
\bibitem [{\citenamefont {Karkheck}\ and\ \citenamefont
  {Stell}(1981)}]{KarkheckStell81}%
  \BibitemOpen
  \bibfield  {author} {\bibinfo {author} {\bibfnamefont {J.}~\bibnamefont
  {Karkheck}}\ and\ \bibinfo {author} {\bibfnamefont {G.}~\bibnamefont
  {Stell}},\ }\href {\doibase 10.1063/1.442154} {\bibfield  {journal} {\bibinfo
   {journal} {J. Chem. Phys.}\ }\textbf {\bibinfo {volume} {75}},\ \bibinfo
  {pages} {1475} (\bibinfo {year} {1981})}\BibitemShut {NoStop}%
\bibitem [{\citenamefont {Stell}\ \emph {et~al.}(1983)\citenamefont {Stell},
  \citenamefont {Karkheck},\ and\ \citenamefont {van
  Beijeren}}]{StellKarkheckVanbeijeren83}%
  \BibitemOpen
  \bibfield  {author} {\bibinfo {author} {\bibfnamefont {G.}~\bibnamefont
  {Stell}}, \bibinfo {author} {\bibfnamefont {J.}~\bibnamefont {Karkheck}}, \
  and\ \bibinfo {author} {\bibfnamefont {H.}~\bibnamefont {van Beijeren}},\
  }\href {\doibase 10.1063/1.446151} {\bibfield  {journal} {\bibinfo  {journal}
  {J. Chem. Phys.}\ }\textbf {\bibinfo {volume} {79}},\ \bibinfo {pages} {3166}
  (\bibinfo {year} {1983})}\BibitemShut {NoStop}%
\bibitem [{\citenamefont {Grmela}\ and\ \citenamefont
  {Garcia-Colin}(1980{\natexlab{a}})}]{GrmelaGarciacolin80}%
  \BibitemOpen
  \bibfield  {author} {\bibinfo {author} {\bibfnamefont {M.}~\bibnamefont
  {Grmela}}\ and\ \bibinfo {author} {\bibfnamefont {L.~S.}\ \bibnamefont
  {Garcia-Colin}},\ }\href {\doibase 10.1103/physreva.22.1295} {\bibfield
  {journal} {\bibinfo  {journal} {Phys. Rev. A}\ }\textbf {\bibinfo {volume}
  {22}},\ \bibinfo {pages} {1295} (\bibinfo {year}
  {1980}{\natexlab{a}})}\BibitemShut {NoStop}%
\bibitem [{\citenamefont {Grmela}\ and\ \citenamefont
  {Garcia-Colin}(1980{\natexlab{b}})}]{GrmelaGarciacolin80b}%
  \BibitemOpen
  \bibfield  {author} {\bibinfo {author} {\bibfnamefont {M.}~\bibnamefont
  {Grmela}}\ and\ \bibinfo {author} {\bibfnamefont {L.~S.}\ \bibnamefont
  {Garcia-Colin}},\ }\href {\doibase 10.1103/physreva.22.1305} {\bibfield
  {journal} {\bibinfo  {journal} {Phys. Rev. A}\ }\textbf {\bibinfo {volume}
  {22}},\ \bibinfo {pages} {1305} (\bibinfo {year}
  {1980}{\natexlab{b}})}\BibitemShut {NoStop}%
\bibitem [{\citenamefont {Grmela}(1981)}]{Grmela81}%
  \BibitemOpen
  \bibfield  {author} {\bibinfo {author} {\bibfnamefont {M.}~\bibnamefont
  {Grmela}},\ }\href@noop {} {\bibfield  {journal} {\bibinfo  {journal} {Can.
  J. Phys.}\ }\textbf {\bibinfo {volume} {59}},\ \bibinfo {pages} {698}
  (\bibinfo {year} {1981})}\BibitemShut {NoStop}%
\bibitem [{\citenamefont {Benilov}\ and\ \citenamefont
  {Benilov}(2018)}]{BenilovBenilov18}%
  \BibitemOpen
  \bibfield  {author} {\bibinfo {author} {\bibfnamefont {E.~S.}\ \bibnamefont
  {Benilov}}\ and\ \bibinfo {author} {\bibfnamefont {M.~S.}\ \bibnamefont
  {Benilov}},\ }\href {\doibase 10.1103/physreve.97.062115} {\bibfield
  {journal} {\bibinfo  {journal} {Phys. Rev. E}\ }\textbf {\bibinfo {volume}
  {97}},\ \bibinfo {pages} {062115} (\bibinfo {year} {2018})}\BibitemShut
  {NoStop}%
\bibitem [{\citenamefont {Linstrom}\ and\ \citenamefont
  {Mallard}(1997)}]{LinstromMallard97}%
  \BibitemOpen
  \bibfield  {author} {\bibinfo {author} {\bibfnamefont {P.~J.}\ \bibnamefont
  {Linstrom}}\ and\ \bibinfo {author} {\bibfnamefont {W.~G.}\ \bibnamefont
  {Mallard}},\ }\href {\doibase 10.18434/t4d303} {\enquote {\bibinfo {title}
  {{NIST} chemistry webbook, {NIST} standard reference database number 69},}\ }
  (\bibinfo {year} {1997})\BibitemShut {NoStop}%
\bibitem [{Note1()}]{Note1}%
  \BibitemOpen
  \bibinfo {note} {In Ref. \cite {BenilovBenilov18}, $E$ was calculated using
  the isobar with the pressure being twice the critical pressure. The
  difference between that result and the present one is less than 0.5\%, which
  proves the robust nature of the employed method for determining
  $E$.}\BibitemShut {Stop}%
\bibitem [{Note2()}]{Note2}%
  \BibitemOpen
  \bibinfo {note} {Note that the EoS-derived calibration proposed in\ Ref.
  \cite {BenilovBenilov18} is sufficiently accurate only for the EoS, but is
  less so for the parameters of the saturated vapor and liquid (which we did
  not realize at the time).}\BibitemShut {Stop}%
\bibitem [{\citenamefont {Tegeler}\ \emph {et~al.}(1999)\citenamefont
  {Tegeler}, \citenamefont {Span},\ and\ \citenamefont
  {Wagner}}]{TegelerSpanWagner99}%
  \BibitemOpen
  \bibfield  {author} {\bibinfo {author} {\bibfnamefont {C.}~\bibnamefont
  {Tegeler}}, \bibinfo {author} {\bibfnamefont {R.}~\bibnamefont {Span}}, \
  and\ \bibinfo {author} {\bibfnamefont {W.}~\bibnamefont {Wagner}},\ }\href
  {\doibase 10.1063/1.556037} {\bibfield  {journal} {\bibinfo  {journal} {J.
  Phys. Chem. Ref. Data}\ }\textbf {\bibinfo {volume} {28}},\ \bibinfo {pages}
  {779} (\bibinfo {year} {1999})}\BibitemShut {NoStop}%
\bibitem [{\citenamefont {Frezzotti}\ and\ \citenamefont
  {Barbante}(2017)}]{FrezzottiBarbante17}%
  \BibitemOpen
  \bibfield  {author} {\bibinfo {author} {\bibfnamefont {A.}~\bibnamefont
  {Frezzotti}}\ and\ \bibinfo {author} {\bibfnamefont {P.}~\bibnamefont
  {Barbante}},\ }\href {\doibase 10.1299/mer.16-00540} {\bibfield  {journal}
  {\bibinfo  {journal} {Mech. Eng. Rev.}\ }\textbf {\bibinfo {volume} {4}},\
  \bibinfo {pages} {16} (\bibinfo {year} {2017})}\BibitemShut {NoStop}%
\bibitem [{\citenamefont {Frezzotti}\ \emph {et~al.}(2018)\citenamefont
  {Frezzotti}, \citenamefont {Gibelli}, \citenamefont {Lockerby},\ and\
  \citenamefont {Sprittles}}]{FrezzottiGibelliLockerbySprittles18}%
  \BibitemOpen
  \bibfield  {author} {\bibinfo {author} {\bibfnamefont {A.}~\bibnamefont
  {Frezzotti}}, \bibinfo {author} {\bibfnamefont {L.}~\bibnamefont {Gibelli}},
  \bibinfo {author} {\bibfnamefont {D.~A.}\ \bibnamefont {Lockerby}}, \ and\
  \bibinfo {author} {\bibfnamefont {J.~E.}\ \bibnamefont {Sprittles}},\ }\href
  {\doibase 10.1103/physrevfluids.3.054001} {\bibfield  {journal} {\bibinfo
  {journal} {Phys. Rev. Fluids}\ }\textbf {\bibinfo {volume} {3}},\ \bibinfo
  {pages} {054001} (\bibinfo {year} {2018})}\BibitemShut {NoStop}%
\end{thebibliography}%

\end{document}